\documentclass[10pt]{article}
\usepackage{fancyhdr}
\usepackage{extramarks}
\usepackage{amsmath}
\usepackage{amsthm}
\usepackage{amsfonts}
\usepackage{siunitx}
\usepackage{tikz}
\usepackage[plain]{algorithm}
\usepackage{algpseudocode}
\usepackage{multirow}
\usepackage{booktabs}
\usepackage{palatino}
\usepackage{graphicx}
\usepackage{subfigure}
\usepackage[colorlinks,linkcolor=black,anchorcolor=black,citecolor=black,urlcolor=blue]{hyperref}
\usepackage{amsmath,bm}
\usepackage{booktabs}
\usepackage{mathtools}
\usepackage{amssymb}
\usepackage{caption}
\usepackage{capt-of}
\usepackage{mciteplus}
\usepackage{cite}
\usepackage{mathrsfs}
\usepackage[title,titletoc,toc]{appendix}
\usepackage{xr}
\usepackage{parskip}
\usepackage{soul}
\usepackage{textcomp}
\usepackage[colaction]{multicol}
\usepackage[switch]{lineno}
\usepackage{lipsum}
\usepackage{etoolbox}
\usepackage{longtable}
\usepackage{array}
\usepackage{tablefootnote}
\newcolumntype{C}[1]{>{\centering\arraybackslash}p{#1}}
\usetikzlibrary{automata,positioning}
\usetikzlibrary{automata,positioning}

\begin{document}

\title{Host immune response driving SARS-CoV-2 evolution}

\author{ Rui Wang$^1$, Yuta Hozumi$^1$, Yong-Hui Zheng$^2$, Changchuan Yin$^3$ \footnote{Address correspondences to Changchuan Yin. E-mail:cyin1@uic.edu}, and Guo-Wei Wei$^{1,4,5}$ \footnote{Address correspondences to Guo-Wei Wei. E-mail:wei@math.msu.edu} \\
$^1$ Department of Mathematics,
Michigan State University, MI 48824, USA\\
$^2$ Department of Microbiology and Molecular Genetics,\\
Michigan State University, MI 48824, USA\\
$^3$ Department of Mathematics, Statistics, and Computer Science, \\
University of Illinois at Chicago, Chicago, IL 60607, USA\\
$^4$  Department of Biochemistry and Molecular Biology\\
Michigan State University, MI 48824, USA \\
$^5$ Department of Electrical and Computer Engineering \\
Michigan State University, MI 48824, USA }

\date{}

\maketitle
\begin{abstract}
The transmission and evolution of severe acute respiratory syndrome coronavirus 2 (SARS-CoV-2) are of paramount importance to the controlling and combating of coronavirus disease 2019 (COVID‑19) pandemic. Currently, near 15,000 SARS-CoV-2 single mutations have been recorded, having a great ramification to the development of diagnostics, vaccines, antibody therapies, and drugs. However, little is known about SARS-CoV-2 evolutionary characteristics and general trend. In this work, we present a comprehensive genotyping analysis of existing SARS-CoV-2 mutations. We reveal that host immune response via APOBEC  and ADAR gene editing gives rise to near 65\% of recorded mutations.  
Additionally, we show that children under age five and the elderly may be at high risk from COVID-19 because of their overreacting to the viral infection.  
Moreover, we uncover that populations of Oceania and Africa react significantly more intensively to SARS-CoV-2 infection than those of  Europe and Asia, which may explain why African Americans were shown to be at increased risk of dying from COVID-19,  in addition to their high risk of getting sick from COVID-19 caused by systemic health and social inequities.  Finally, our study indicates that for two viral genome sequences of the same origin, their evolution order may be determined from the ratio of mutation type C$>$T over T$>$C. 

\end{abstract}

\tableofcontents

\newpage

\section{Introduction}
The ongoing raging outbreak of coronavirus disease 2019 (COVID‑19) caused by severe acute respiratory syndrome coronavirus 2 (SARS-CoV-2) has led to tremendous human mortality and economic hardship. As of July 31, 2020, over 17106007 confirmed COVID-19 cases had been reported worldwide and 668910 deaths have occurred from the disease \cite{who_2020}. To mitigate this devastating pandemic, we have to control its spread by sufficient testing, social distancing, contact tracking, and developing effective diagnosis tools, efficacious antiviral drugs, antibody therapies, and preventive vaccines.

SARS-CoV-2 is a positive-sense single-strand RNA virus that belongs to the beta coronavirus genus \cite{wu2020new}. It has a genome size of 29.82 kb, which encodes multiple non-structural and structural proteins. The leader sequence and ORF1ab encode non-structural proteins for RNA replication and transcriptions. The downstream regions of the genome encode structural proteins, including the spike (S) protein, the nucleocapsid (N) protein, the envelope (E) protein, and the membrane (M) protein. All of the four major structural proteins are required to produce a structurally complete viral particle. The S protein mediates viral attachment to host angiotensin-converting enzyme 2 (ACE2)  receptor and subsequent fusion between the viral and host cell membranes aided by transmembrane serine protease 2 (TMPRSS2) to allow the entry of viruses into the host cell \cite{xiao2003sars,glowacka2011evidence,hoffmann2020sars}. The nucleocapsid (N) protein, one of the most abundant viral proteins, binds to the RNA genome and is involved in replication processes, assembly, and host cellular response during viral infection \cite{mcbride2014coronavirus}.

Mutagenesis is a basic biological process that changes the genetic information of organisms.  As a primary source for many kinds of cancer and heritable diseases, mutagenesis maybe fearful but is a driving force for natural evolution \cite{yue2005loss,stefl2013molecular}. Although viruses are not organisms per se, they are at the edge of life. Our \href{https://users.math.msu.edu/users/weig/SARS-CoV-2_Mutation_Tracker.html}{SARS-CoV-2 Mutation Tracker}  (\url{https://users.math.msu.edu/users/weig/SARS-CoV-2_Mutation_Tracker.html}) shows that near 15,000 mutations have occurred on SARS-CoV-2 \cite{wang2020mutations}. More than 1000 mutations on the S protein gene have a significant impact on SARS-CoV-2 infectivity  \cite{korber2020tracking,grubaugh2020making,chen2020mutations}.  These mutations should be put into the perspective that COVID-19 has globally spread. The geographical and demographical diversity of the viral transmission and exogenous and endogenous genotoxins exposures have stimulated SARS-CoV-2 mutations. If we consider the average number of mutations per genome, SARS-CoV-2 is mutating slower than other viruses, such as the flu and  common cold viruses. This is because SARS-CoV-2 belongs to the coronaviridae family and the Nidovirales order, which has a genetic proofreading mechanism in its replication achieved by an enzyme called non-structure protein 14 (NSP14) in synergy with NSP12, i.e.,  RNA-dependent RNA polymerase (RdRp) \cite{sevajol2014insights, ferron2018structural}. As a result, SARS-CoV-2 has a relatively high fidelity in its transcription and replication process.  In general, Coronavirus mutations are created from three major sources, namely, random errors in replication, such as genetic drift and spontaneous genotoxins,  viral replication proofreading and defective repair mechanisms,  and host immune responses, such as destructive gene editing \cite{sanjuan2016mechanisms, grubaugh2020making}.  Genotyping tracks mutations overpopulation, space, and time, while also providing a method to understand the molecular mechanism of SARS-CoV-2 proteins, protein-protein interactions,  and their synergy with host cell proteins, enzymes, and signaling pathways. 

The studies of SARS-CoV genomes have so far predominantly focused on understanding genome mutation variants,   implications in virus transmissions \cite{yin2020genotyping,phan2020genetic}, and ramifications on the development of diagnostics \cite{wang2020mutations,khan2020presence}, vaccines \cite{wang2020decoding}, antibodies \cite{baum2020antibody}, and drugs \cite{wang2020decoding}. 
 
Although it is difficult to determine the detailed mechanism of every specific mutation, early work on a few initial SARS-CoV-2 strains in Wuhan, China, revealed that hypermutations C$>$T are most likely resulted from the APOBEC (apolipoprotein B mRNA editing enzyme, catalytic polypeptide-like) deamination in RNA editing \cite{di2020evidence}. In the standard genetic code, all three stop codons, TAA, TAG, and TGA, involve T but not C. Therefore, the gene-editing imposed C$>$T mutations will have a high possibility to terminate the translation of viral proteins, which undermines viral functions and survivability. Both spontaneous C$>$T transitions and APOBEC   deamination are regarded as genotoxins and can lead to cancers for humans. There are two well-known deaminase RNA editing mechanisms in human cells: the APOBEC  \cite{zheng2004human} and the ADAR (adenosine deaminases acting on RNA) \cite{nishikura2016editing}. The APOBEC enzymes deaminate cytosines into uracils (C$>$U) on single-stranded nucleic acids (ssDNA or ssRNA). It is well established that the human genome encodes activation-induced cytidine deaminases (AIDs) and several homologous APOBEC cytidine deaminases that function in innate immunity as well as in RNA editing \cite{smith2012functions,liu2018aid}. In both innate and adaptive immunity, AID and APOBEC cytidine deaminases modulate immune responses by mutating specific nucleic acid sequences of hosts and pathogens. The ADAR enzymes deaminate adenines into inosines (A-to-I) and result in A$>$G mutation. The significance of A-to-I editing is appreciated for its abundance in both host and viral RNAs. ADAR enzymes play important roles during viral infections. They can have either a proviral or an antiviral consequence, dependent upon the virus-host combination \cite{samuel2011adenosine,gonzales2017making}. 

The APOBEC family proteins play critical functional roles within the adaptive and innate immune system, which involves at early times after the infection \cite{harris2015apobecs}. Therefore, the higher ratio of C$>$T mutations may indicate the strong capacity of the host immune system. However, a strong immune response is a double-edged sword. On the one hand, it may help host cells to defeat the virus more efficiently. On the other hand, it can result in a ``cytokine storm”, which is a key cause of the death of  COVID-19 patients by the exponential growth of inflammation and organ damage \cite{song2020cytokine}. 

In this work, we analyze a large volume of single nucleotide polymorphisms (SNPs) found in 33693 complete SARS-CoV-2 genome isolates globally. By analyzing the distribution of 12 SNP types, we notice that the ratio of C$>$T mutations is predominately higher than that of the other types of mutations, indicating that hypermutation C$>$T may result from extensive host RNA editing, i.e.,  the APOBEC deamination. 
Additionally, we investigate the distribution of 12 SNP types in different age groups, gender groups, and geographic locations to understand whether these hypermutations have the age/gender/demographic preference. 
Moreover, we provide deep insights into the mutation motif and hot-spot patterns from 13833 single mutations decoded from 33693 complete SARS-CoV-2 genome sequences, revealing mutational signatures and preferred genetic environments. 
Finally, we hypothesize that virus genomes evolve through host innate immune response imposed gene editing,  i.e.,  C$>$T, and virus protective mechanism-installed defective revisionary mutations, T$>$C. As a result, both C$>$T and T$>$C mutation ratios are usually high. We show that the ratio of C$>$T to T$>$C mutations is higher than the unity in the forward viral evolution, which suggests the master and slave relationship between host gene editing and virus protective mechanism. Therefore,  we propose the use of the C$>$T to T$>$C ratio being higher than the unity ($>$1) as the indication of the forward viral evolution direction. 

\section{Results}
To reveal that C$>$T and A$>$G mutations are driven by RNA-APOBEC and RNA-ADAR editing, we first analyze 33693 complete SARS-CoV-2 genome sequences and total 13833 single mutations are found as of July 31, 2020. To be noted, 13833 single mutations are unique mutations, i.e., the same mutation that appears in different SARS-CoV-2 isolates is only counted once. If we count the same mutation in different SARS-CoV-2 isolates repeatedly according to their frequency, then all of the mutations that are detected in the 33693 complete SARS-CoV-2 genome sequences are called  non-unique mutations. With the reference sequence of SARS-CoV-2 genome collected on January 5, 2020 \cite{wu2020new},  we calculate the proportion of 12 SNP types (i.e., A$>$T, A$>$C, A$>$G, T$>$A, T$>$C, T$>$G, C$>$T, C$>$A, C$>$G, G$>$T, G$>$C, G$>$A) worldwide.  The unusually high ratios of C$>$T and A$>$G mutations indicate that RNA-APOBEC editing and RNA-ADAR editing are involved in the host immune response to SARS-CoV-2 infection. Additionally,  to understand gene-editing preference, we investigate the distribution of 12 SNP types of mutations in different countries/regions, age groups, and gender groups. Furthermore, we decode mutation motifs from the 2-mer and 3-mer sequence contexts to survey the hot-spot patterns and mutational signatures driven by gene-editing. Moreover, we analyze the proportion of 12 SNP types among SARS-CoV, Bat-SL-BM48-31, Bat-SL-CoVZC45, Bat-SL-RaTG13, and SARS-CoV-2. We discover that the viral evolution order can be determined by the ratios of C$>$T/T$>$C. These results are presented in following subsections.

\subsection{Host immune response to SARS-CoV-2 infection with gene editing }
\subsubsection{Global  analysis}\label{sec:global}
 
\autoref{tab:world ratio} illustrates the proportion of 12 SNP types of SARS-CoV-2 (i.e., A$>$T, A$>$C, A$>$G, T$>$A, T$>$C, T$>$G, C$>$T, C$>$A, C$>$G, G$>$T, G$>$C, G$>$A) in the global. Here we only consider the unique SNPs. 

\begin{table}[ht!]
    \centering
    \setlength\tabcolsep{3pt}
	\captionsetup{margin=0.1cm}
	\caption{The distribution of 12 SNP types among unique mutations in the SARS-CoV-2 genome isolates worldwide. The unique SNP mutations are considered in the calculation, i,e., the same type of mutations in different genome isolates is only counted once.}
    \label{tab:world ratio}
    \begin{tabular}{cccccccccccccc}
    \hline
    Type & A$>$T & A$>$C & A$>$G & T$>$A & T$>$C & T$>$G & C$>$T & C$>$A & C$>$G & G$>$T & G$>$C & G$>$A\\
    \hline
    Ratio & 4.44\% & 3.75\% & 14.87\% & 3.43\% & 14.53\% & 2.80\% & 24.06\% & 4.00\% & 1.25\% & 13.33\% & 2.36\% & 11.17\% \\
    \hline
    \end{tabular}
\end{table}

\begin{figure}[ht!]
    \centering
    \includegraphics[width=1.1\textwidth]{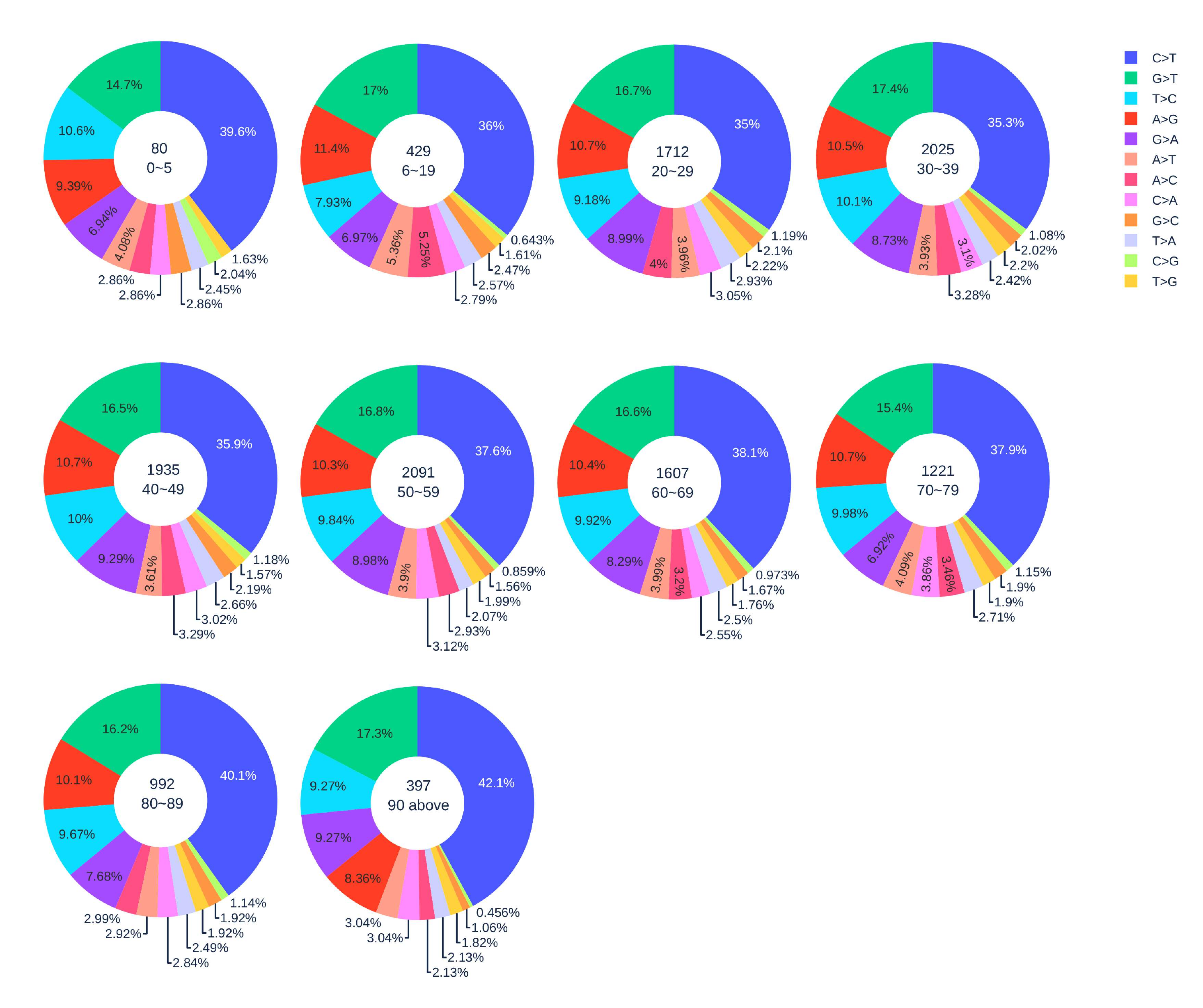}
    \caption{The distribution of 12 SNP types among unique mutations in the SARS-CoV-2 genome isolates from different age groups. The text inside each circle represents for the total number of records that have the age information in different age groups.}
    \label{fig:Age_World}
\end{figure}

 First, it is noticed that not all SARS-CoV-2 mutations are created equal. Mutation C$>$G only accounts for 1.25\%. A few other mutation types, G$>$C, T$>$G, T$>$A, and A$>$C, are not frequent either.  If mutations are random, each mutation should have a ratio of 8.3\% on average. It can be seen that C$>$T owns the largest proportion (24.06\%), which is much higher than the average ratio. Therefore,  the hypermutation C$>$T must be driven by additional mechanisms. It is all known that host RNA-APOBEC editing leads to excessive  C$>$T transitions.

Moreover, the second most frequent mutation type  is A$>$G transition. Its ratio of  14.87\%  A$>$G is much higher than the average ratio of 8.3\%, indicating that RNA-ADAR editing is also involved in the host immune response. Although the high ratios of C$>$T and A$>$G reveal that the immune system is combating with SARS-CoV-2 by two deaminase RNA editing mechanisms, the relatively high ratios of the reversed mutations T$>$C and G$>$A also indicate that SARS-CoV-2 fights back the destructive gene editing using its defective proofreading and repairing mechanisms.

\begin{figure}[ht!]
    \centering
    \includegraphics[width=0.7\textwidth]{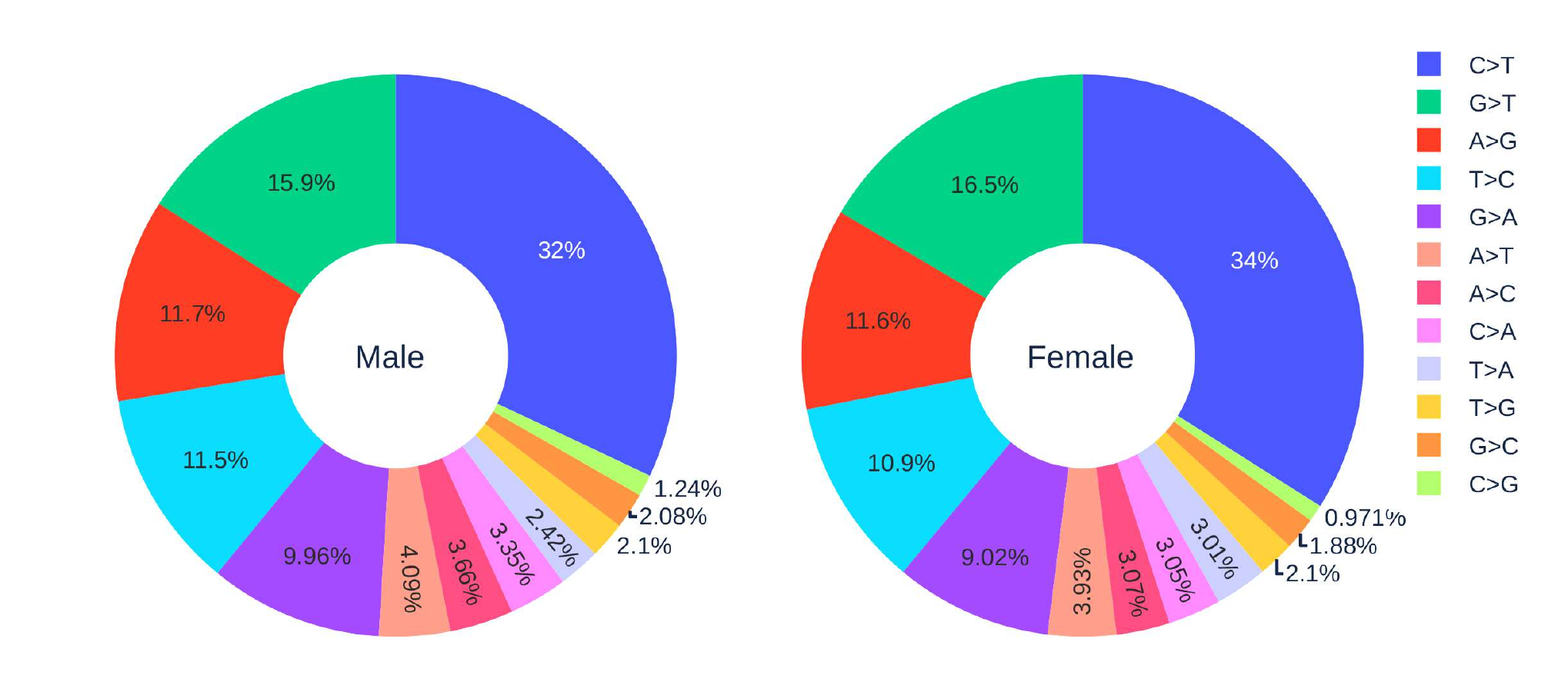}
    \caption{The distribution of 12 SNP types among unique mutations in the SARS-CoV-2 genome isolates from two gender groups. The text inside each circle represents for the total number of records that have the gender information in different gender groups.}
    \label{fig:Gender_World}
\end{figure}

Finally, it is well-known that mutations can be classified into four transition types  (i.e., A$>$G,  G$>$A, C$>$T, and T$>$C) and eight transversion types.  \autoref{tab:world ratio} shows that all transition types have relatively high ratios. Whereas, all transversion types, except for G$>$T, have relatively low ratios.  This is due to the fact that it is easier to substitute a single ring nucleotide structure for another single ring nucleotide structure than to substitute a double ring nucleotide for a single ring nucleotide. Additionally,  transitions are more likely to result in silent mutations. Therefore,  transversions can be more destructive to viral genomes.

\subsubsection{Age analysis}

\begin{figure}[ht!]
    \centering
    \includegraphics[width=1\textwidth]{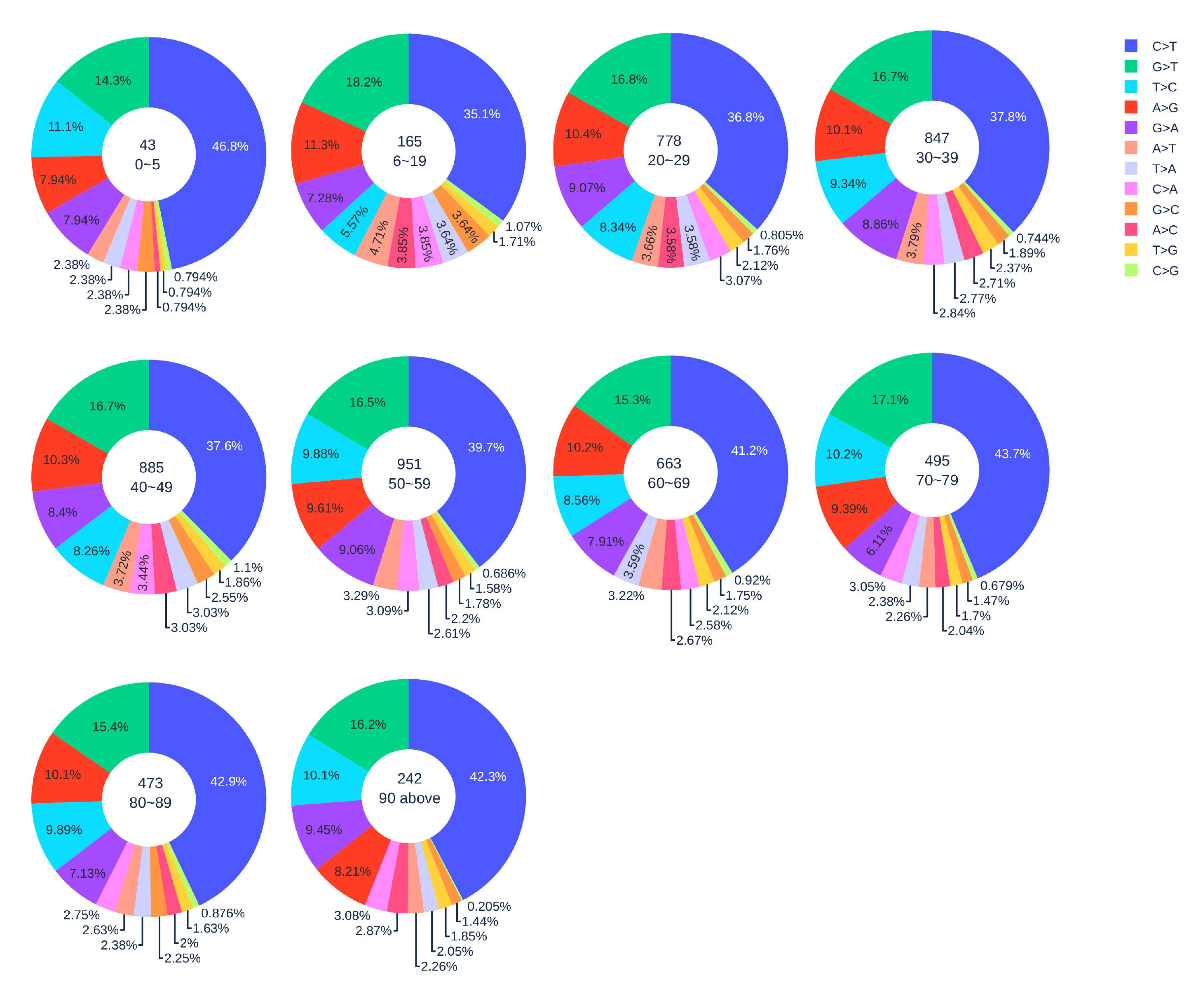}
    \caption{The distribution of 12 SNP types among unique mutations in the SARS-CoV-2 genome isolates from different age groups among female patients. The text inside each circle represents for the total number of records that have the age information in different age groups.}
    \label{fig:Age_Female_World}
\end{figure}

\autoref{fig:Age_World} illustrates the distribution of 12 SNP types among unique SNPs in SASR-CoV-2 genome isolates from different age groups. In general, with the increase of age, the ratio of C$>$T gradually increased. Here, 42.1\% C$>$T mutations are detected in patients who are older than 90 years old, indicating that the immune systems in elderly patients may fight against the SARS-CoV-2 harder than the immune systems in young patients. However, the severe COVID-19 cases may be due to immune systems' over response. When SARS-CoV-2 infects a host cell, a set of proteins called cytokines will be released from a broad range of cells (mainly immune cells). Cytokines are involved in the immune response to produce more immune cells and recruit them to the sites of inflammation in order to fight against the viral infection. In turn, more cytokines can be released from the immune cells. This positive feedback loop will result in a ``cytokine storm", which can beget the exponential growth of inflammation, trigger apoptosis, and lead to organ damage \cite{song2020cytokine}. Therefore, we hypothesize that if the immune system overreacts to the invading pathogens, it is more likely to cause the cytokine storm and aggravate the condition of the COVID-19 patients. It can be seen in \autoref{fig:Age_World}, patients who are older than 80 years old have more C$>$T mutations compared to other age groups. This result reveals that the APOBEC3 activity in the immune system is more active and the immune response is stronger in older people. Consequently, the cytokine storm may happen more frequently in older people  than it does in younger people. This might be one of the main causes of the high COVID-19 fatality for the elderly. Age-related mutagenesis, i.e.,  C$>$T transition,  is known to cause more cancers diagnostics in the elderly \cite{alexandrov2013signatures}. 
\begin{figure}[ht!]
    \centering
    \includegraphics[width=1\textwidth]{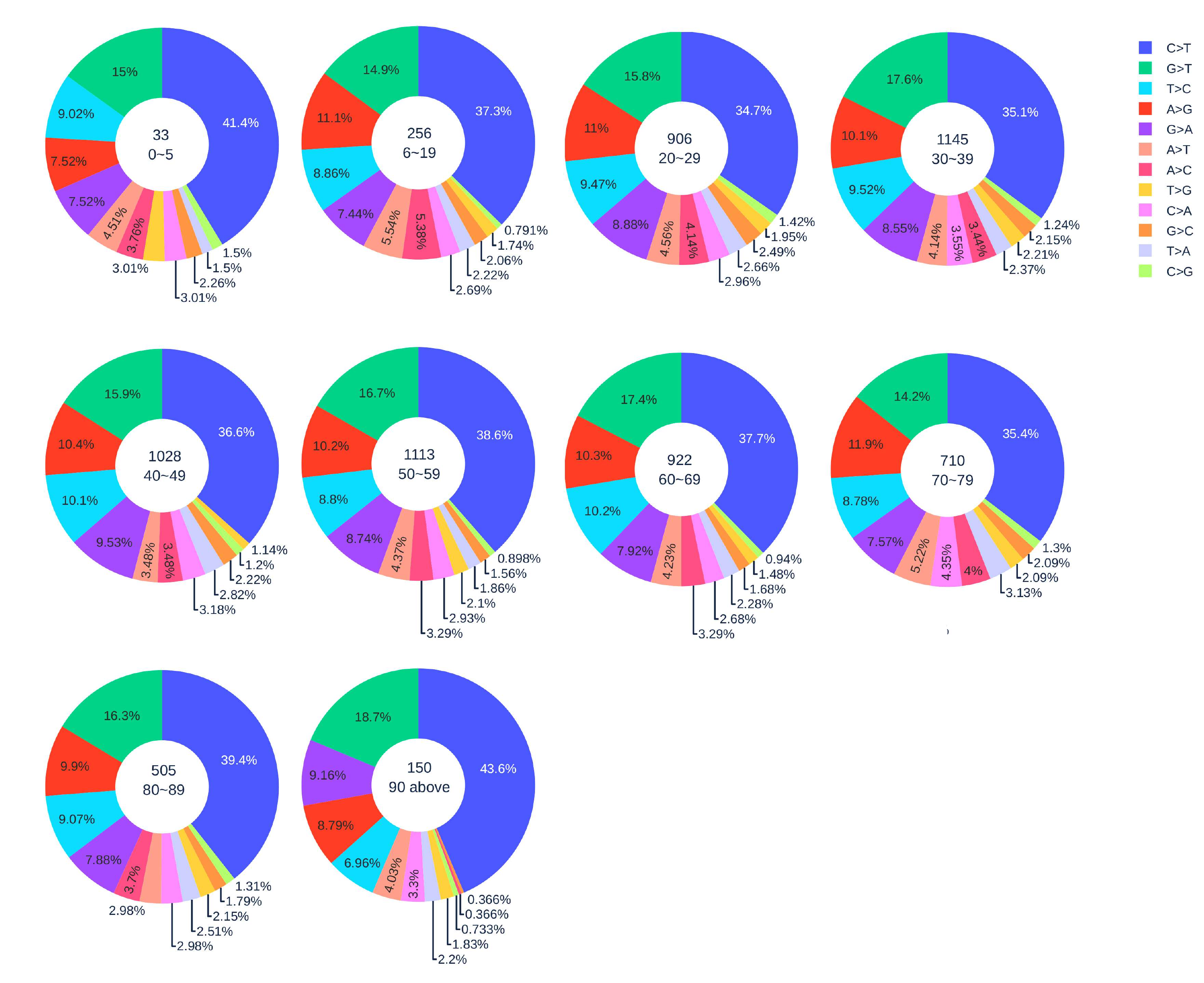}
    \caption{The distribution of 12 SNP types among unique mutations in the SARS-CoV-2 genome isolates from different age groups among male patients. The text inside each circle represents for the total number of records that have the age information in different age groups.}
    \label{fig:Age_Male_World}
\end{figure}

Notably, the SARS-CoV-2 samples from children under five years old have a relatively high ratio of C$>$T mutations (39.6\%), indicating that they also have a relatively active immune response when fighting against SARS-CoV-2. Moreover, the reversed mutation type T$>$C for samples from children under five years old and adults older than 90 years old has the second-largest ratio. 
In other age groups, T$>$C has the fourth-largest ratio. As demonstrated before, the reversed mutation T$>$C may reveal that SARS-CoV-2 is capable of fighting back against the host immune system. Therefore, we deduce that SARS-CoV-2 will fiercely counter-attack against the immune system in children under five and adults older than 90.

Our result reveals that the immune systems of children under five years old are less well-developed and weaker than those of adults. They have to fight more intensively when SARS-CoV-2 infects. This result suggests children under five are at risk of COVID-19. However, the long-term health consequence of young children's unusual response to SARS-CoV-2 infection is to be further studied.

\subsubsection{Gender analysis}

\autoref{fig:Gender_World} shows the distribution of 12 SNP types in SARS-CoV-2 genome isolates globally from two gender groups. The ratio of C$>$T mutations in females is slightly higher in males, which matches the finding that women have a  stronger immune response than men \cite{hewagama2009stronger,klein2012sex}. Moreover, \autoref{fig:Age_Female_World} and \autoref{fig:Age_Male_World} depict the distribution of 12 SNP types in different age groups among female and male patients. Overall, the proportion of C$>$T mutations in the SARS-CoV-2 genomes from females is higher than the C$>$T proportion in the SARS-CoV-2 genomes from male except for the age between 6-19 and older than 90. Therefore, we can deduce that the RNA editing has age and gender preference, it is more likely to happen or become stronger for the females who are older than 90 years old or under 5 years old.

\begin{table}[ht!]
    \centering
    \setlength\tabcolsep{11pt}
	\captionsetup{margin=0.1cm}
	\caption{The number of complete SARS-CoV-2 genomes with age/gender information in the United Kingdom, United States, Australia, India, and the world. }
    \label{tab:countries}
    \begin{tabular}{lcccccccccc}
    \hline
    Country & Total counts & Age counts & gender counts\\
    \hline
    United Kingdom & 10740 & 2159 & 2134\\
    United States & 8729 & 1888 & 2095\\
    Australia & 1329 & 776 & 750\\
    India & 1088 & 1068 & 1071\\
		World &33693 & 12513& 12181\\
    \hline
    \end{tabular}
\end{table}

\begin{figure}[ht!]
    \centering
    \includegraphics[width=1\textwidth]{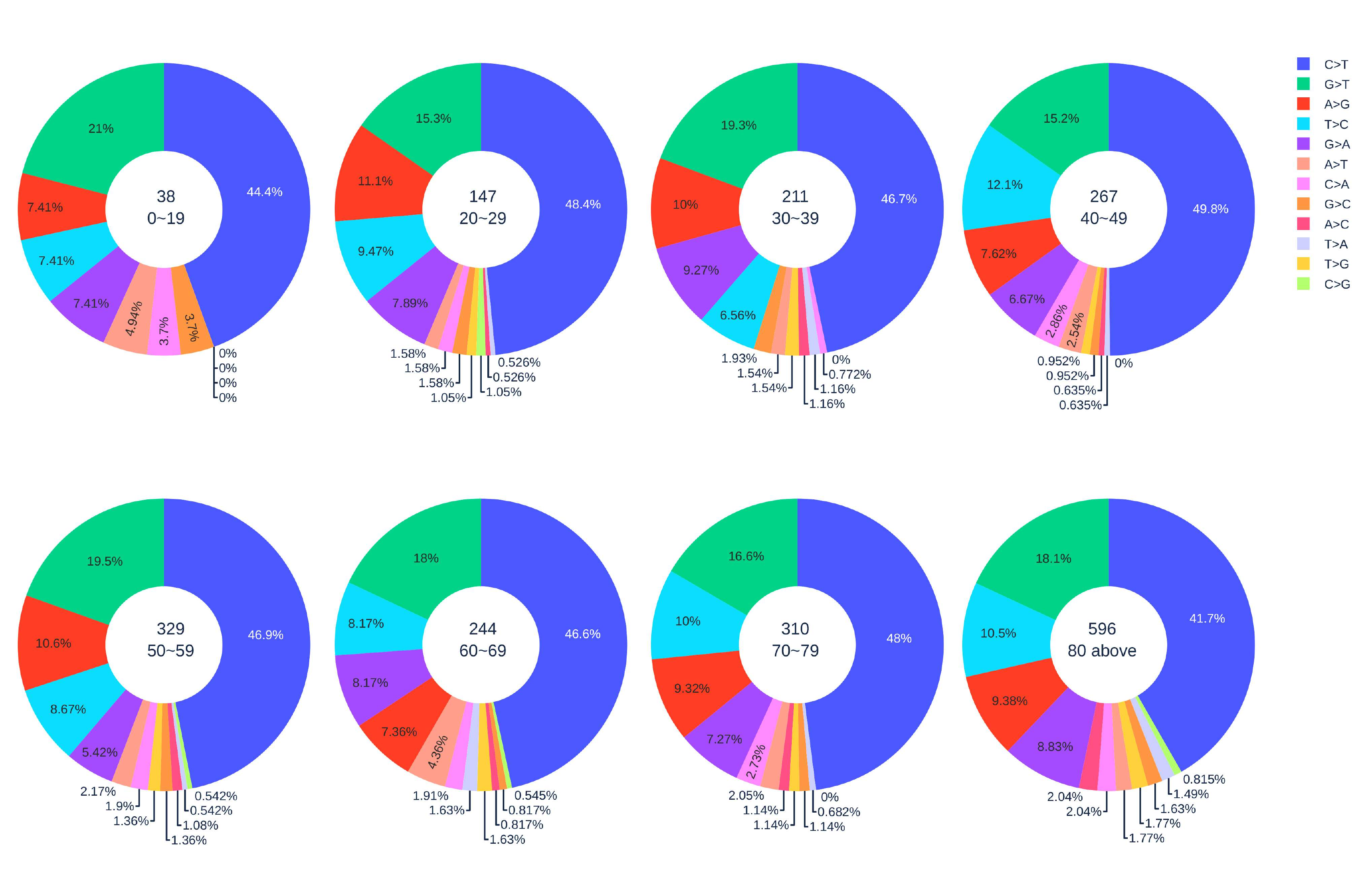}
    \caption{The distribution of 12 SNP types among unique mutations in the SARS-CoV-2 genome isolates from the United Kingdom. The text inside each circle represents for the total number of records in different age groups.}
    \label{fig:UK_Age_World}
\end{figure}

\begin{figure}[ht!]
    \centering
    \includegraphics[width=1\textwidth]{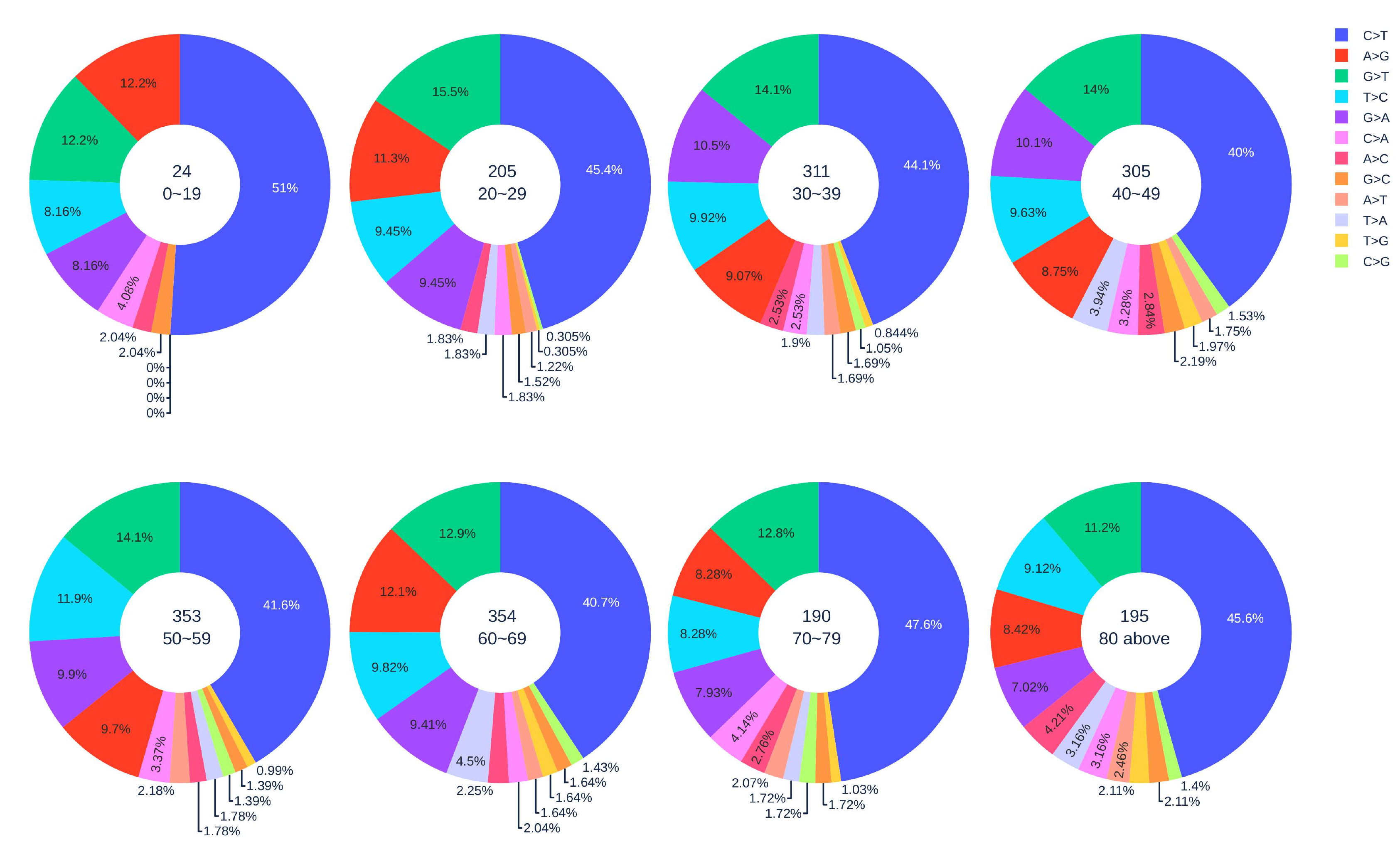}
    \caption{The distribution of 12 SNP types among unique mutations in the SARS-CoV-2 genome isolates from the United States. The text inside each circle represents for the total number of records in different age groups.}
    \label{fig:US_Age_World}
\end{figure}

\subsubsection{Geographic analysis  }

In this section, we analyze the distribution of SARS-CoV-2 mutations in different countries and regions. Limited by the number of complete genome sequences submitted to GISAID that have appropriate labels, we only analyze the countries with more than 1000 labeled sequences to maintain statistical significance.  \autoref{tab:countries} lists the total number of SARS-CoV-2 sequences in the United Kingdom, United States, Australia, and India. The number of sequences with age and gender information is given in \autoref{tab:countries}. 

\begin{figure}[ht!]
    \centering
    \includegraphics[width=1\textwidth]{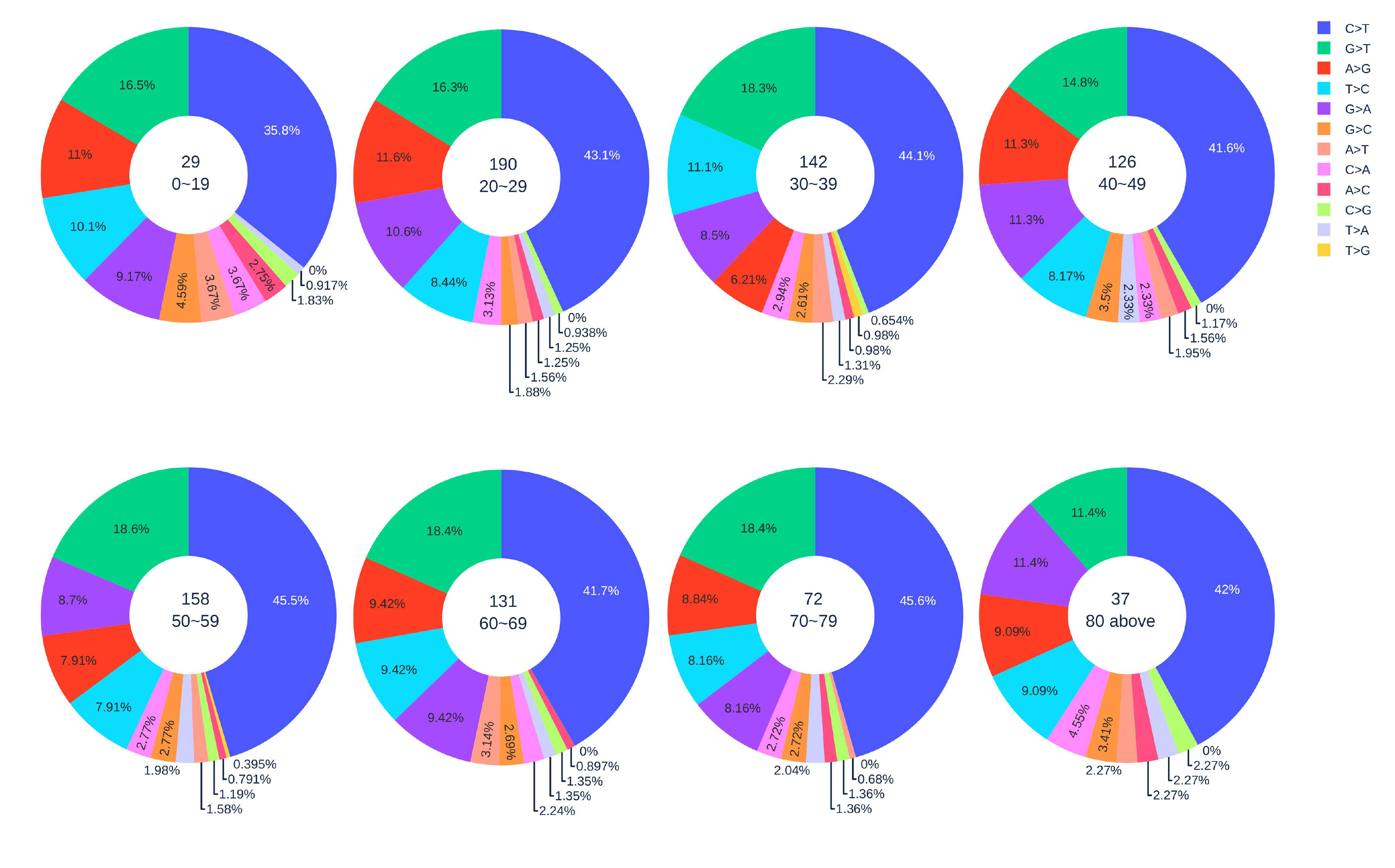}
    \caption{The distribution of 12 SNP types among unique mutations in the SARS-CoV-2 genome isolates from Australia. The text inside each circle represents for the total number of records in different age groups.}
    \label{fig:AU_Age_World}
\end{figure}

\begin{figure}[ht!]
    \centering
    \includegraphics[width=1\textwidth]{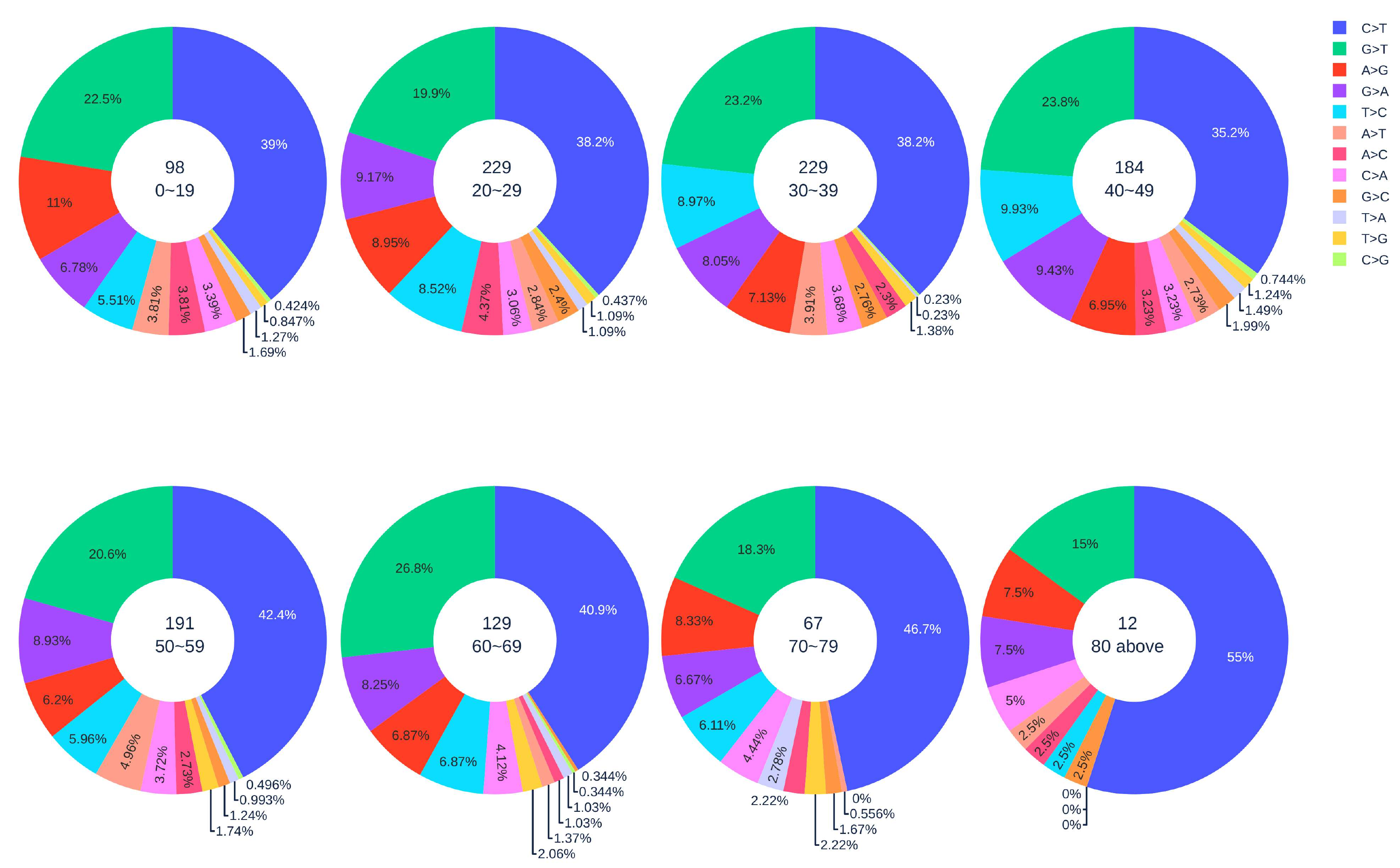}
    \caption{The distribution of 12 SNP types among unique mutations in the SARS-CoV-2 genome isolates from India. The text inside each circle represents for the total number of records in different age groups.}
    \label{fig:IN_Age_World}
\end{figure}

\autoref{fig:UK_Age_World}, \autoref{fig:US_Age_World}, \autoref{fig:AU_Age_World}, and \autoref{fig:IN_Age_World} illustrate the distribution of 12 SNP types in the SARS-CoV-2 genome isolates from different age groups in the United Kingdom, United States, Australia, and India, respectively. We can see that the SARS-CoV-2 genome isolates from the United Kingdom patients have the highest ratio of C$>$T compared to those from the other three countries. It is interesting to note that the SARS-CoV-2 genome isolates from the patients older than 80 years old from the United Kingdom and Australia have less C$>$T mutations, which is not consistent with the global pattern.  

\begin{figure}[ht!]
    \centering
    \includegraphics[width=1\textwidth]{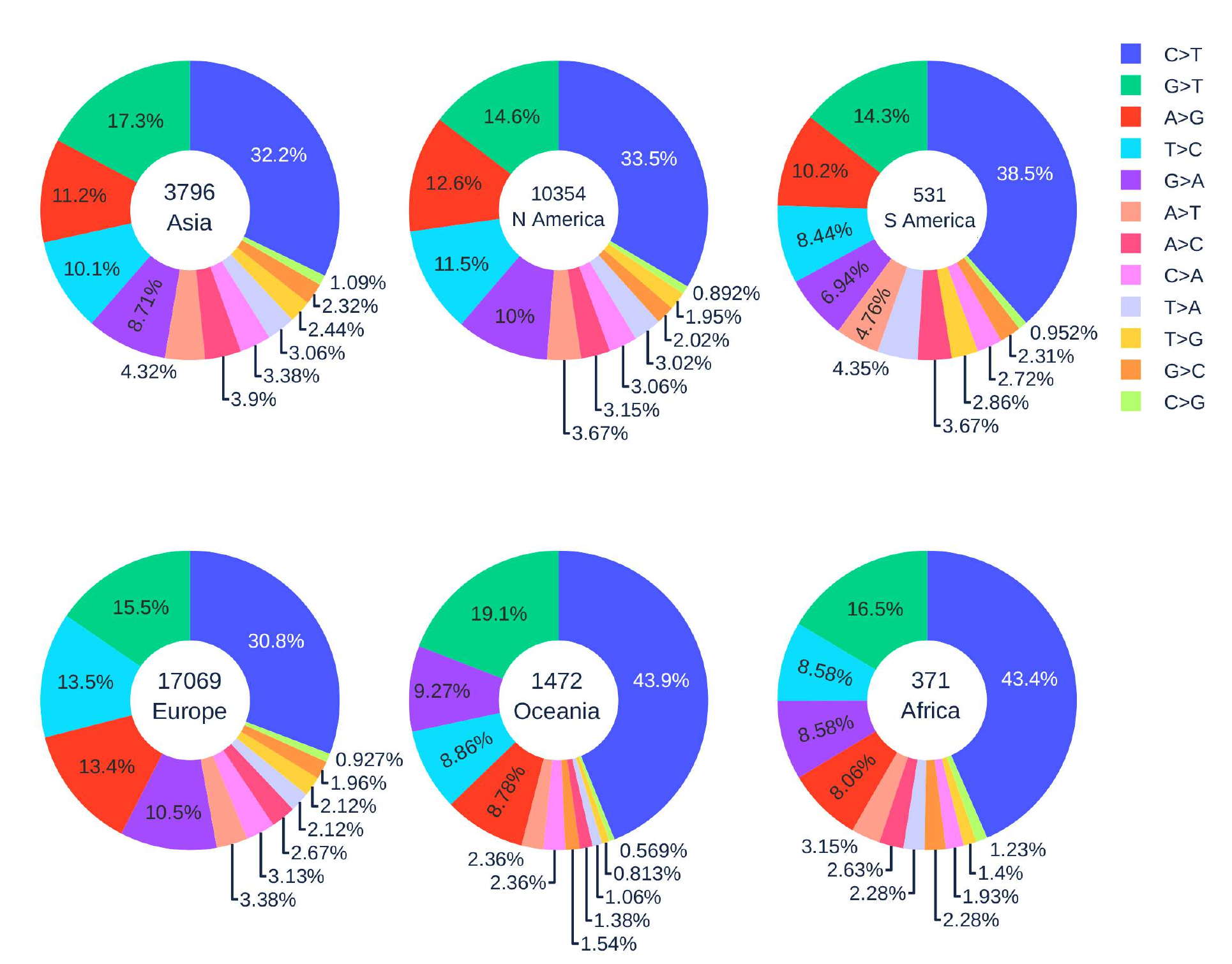}
    \caption{The distribution of 12 SNP types among unique mutations in the SARS-CoV-2 genome isolates in six continents. The text inside each circle represents for the total number of records in each continent.}
    \label{fig:Region}
\end{figure}

\autoref{fig:Region} illustrates the distribution of 12 SNP types  in six continents. The SARS-CoV-2 genome isolates from Europe, Asia, and North America patients have a relatively low C$>$T mutation ratio (less than 35\%), while the reversed T$>$C mutation ratio is relatively high (greater than 10\%). On the contrary, South America, Oceania, and Africa have higher C$>$T ratios but lower T$>$C ratio. It worth noting that the C$>$T mutation ratios in the SARS-CoV-2 genome isolates from Oceania and Africa are more than 10\% higher than those of Asia, Europe, and North America. This result indicates that the APOBEC editing may be more active, and the counterattack of SARS-CoV-2 might be weakened by the strong immune response in the populations of Oceania and Africa.

African Americans, as an ethnic group of Americans with total or partial ancestry from Africa, are genetically associated with Africans. There have been many concerns about the fact that they are disproportionately affected by COVID-19 (\url{https://www.cdc.gov/coronavirus/2019-ncov/community/health-equity/race-ethnicity.html}). The present finding indicates that the immune systems of African Americans may also 
overreact to  SARS-CoV-2 infection by excessive gene editing. 

Another interesting issue is that A$>$G mutation ratios of genome isolates from  Oceania and Africa are very low ($<$9\%). In contrast,  A$>$G mutation ratios of genome isolates from other regions are significantly higher ($>$ 11\%). These results indicate that Asia and Europe populations may have adopted significantly different genetic and molecular mechanisms in their immune response to viral infection compared to those of  Oceania and Africa.  Further studies are required to fully understand these differences.  

\subsection{The SNP preferences on sequence contexts}
 
The mutation preferences in sequence contexts may be used to predict the mutational signatures from genome sequences. Despite numerous studies of the mutation contexts in APOBEC editing inhuman cells, little is known for the mutation contexts in the SARS-CoV-2 genome. As we have a large number of SNP mutations from SARS-CoV-2 genomes, here we discuss the mutation frequencies from 2-mer and 3-mer sequence contexts. We present  4-mer   sequence contexts in the Supporting Information.

In general, the patterns discussed in this section are consistent with those presented in Section \ref{sec:global}. However, this section offers more detailed information about mutational signatures. 

\begin{figure}[ht!]
    \centering
    \includegraphics[width=1\textwidth]{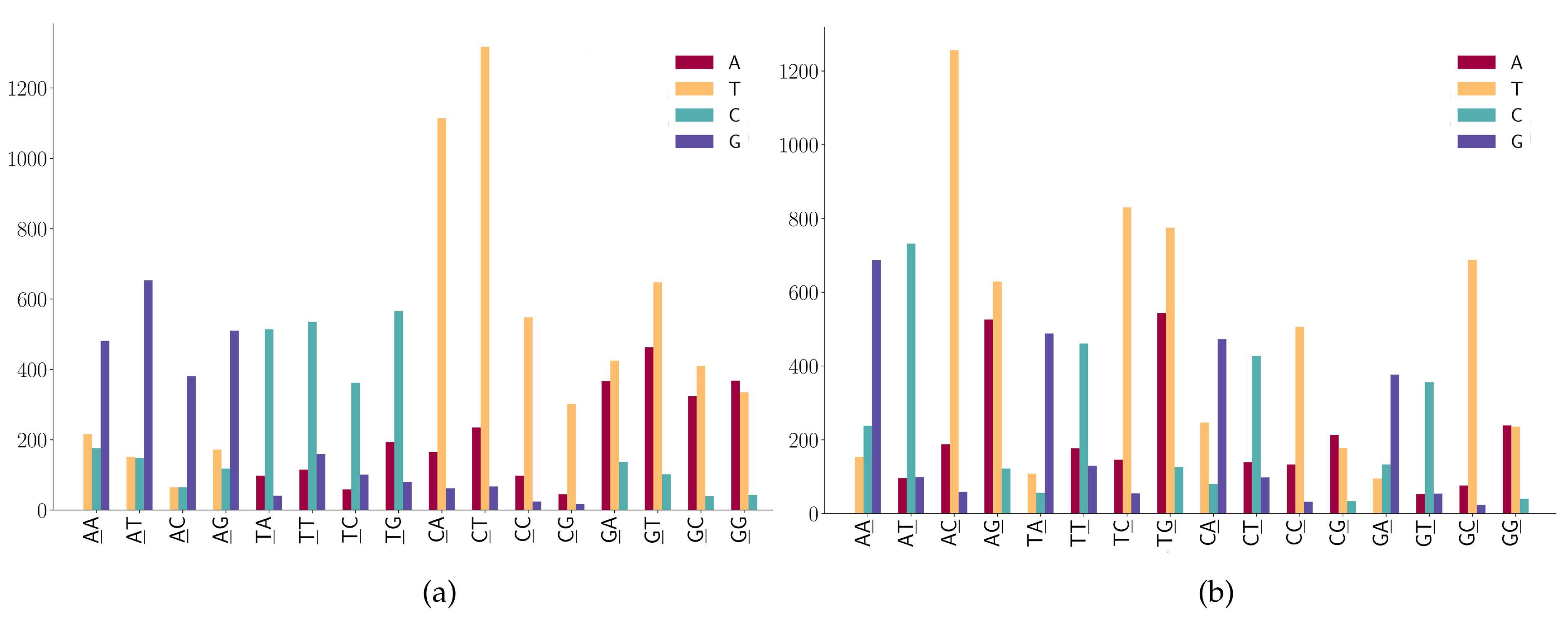}
    \caption{SNP frequencies on 2-mer motifs. (a) SNP frequencies are on the first position of 2-mer motifs. (b) SNP frequencies are on the second position of 2-mer motifs.}
    \label{fig:2mer}
\end{figure}

For mutation motifs of  SNPs at the first position of  2-mers \autoref{fig:2mer}(a), we observe that  motif 2-mer \underline{C}W (where W is either A or T) for C$>$T mutation is the predominant context. Similarly, for mutation motifs of the  SNPs at the second position of  2-mers \autoref{fig:2mer}(b),  motif 2-mer W\underline{C} for C$>$T mutation is the predominant context. These results are consistent with the previous study that T\underline{C}W contexts (where W $=$ A or T) are predominantly caused by APOBEC-catalyzed deamination of cytosine (C) to thymine (T) or uracil (U) in human cancer cells \cite{roberts2013apobec}.

\begin{figure}[ht!]
    \centering
    \includegraphics[width=1\textwidth]{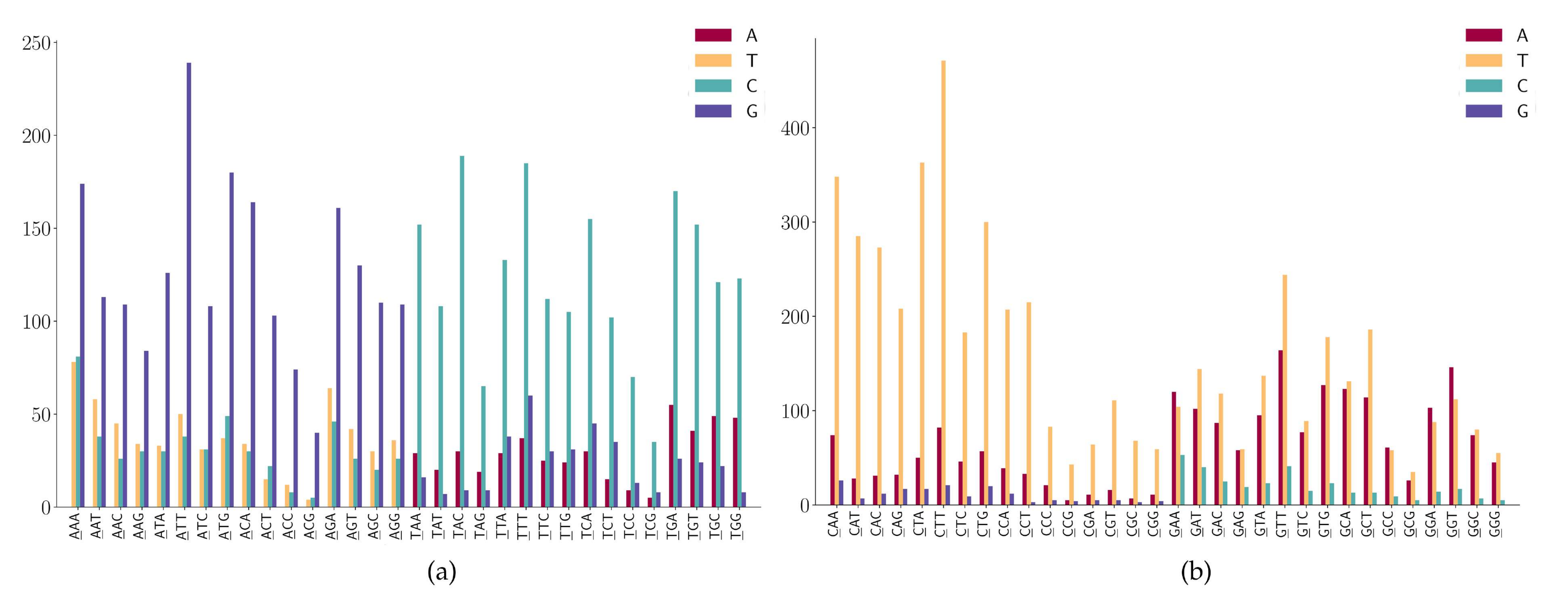}
    \caption{SNP frequency at the first positions of the 3-mer motifs. (a) A or T is at the first positions of 3-mer motifs. (b) C or G is at the first positions of 3-mer motifs.}
    \label{fig:3mer}
\end{figure}

For the SNPs at the first position of  3-mers (\underline{A}NN or \underline{T}NN) (\autoref{fig:3mer}(a)), we observe the following mutation patterns.\\
(1) \underline{A}NN (except for \underline{A}AC and \underline{A}CC) has high A$>$G mutation. \underline{A}AC and \underline{A}CC contexts have a high frequency in A$>$T mutations.\\
(2) \underline{T}NN has a high frequency in T$>$C mutations.\\

For the SNPs at the first position of  3-mers (\underline{C}NN or \underline{G}NN)  shown in \autoref{fig:3mer} (b), we observe the following mutation patterns.\\
(1) \underline{C}NN has  a high frequency in C$>$T mutations. \\
(2) \underline{G}GA has  a high frequency in G$>$C mutations.\\
(3) \underline{G}CA has a relatively   high frequency in G$>$A mutations.\\
(3) \underline{G}GN (N$\neq$A) has relatively high frequency in G$>$A mutations.\\

\begin{figure}[ht!]
    \centering
    \includegraphics[width=1\textwidth]{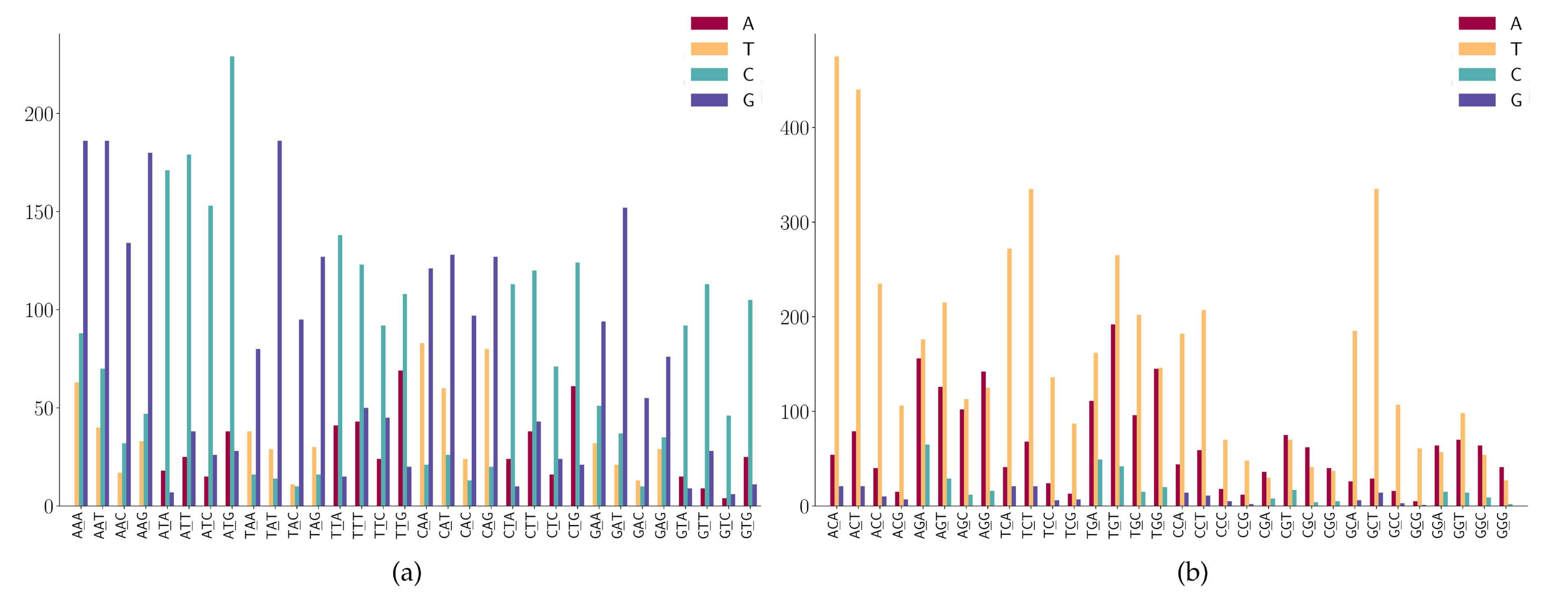}
    \caption{SNP frequency on the second position of  3-mer motifs. (a) A or T is on the second position of 3-mer motifs. (b) C or G is on the second position of 3-mer motifs.}
    \label{fig:3mer_2}
\end{figure}

For the SNPs at the second position of the 3-mers (N\underline{A}N or N\underline{T}N)  as shown in \autoref{fig:3mer_2} (a), we observe the following mutation patterns.\\
(1) N\underline{A}N has a high frequency in A$>$G mutations \\
(2) N\underline{T}N has a high frequency in T$>$C mutations \\
(3) The T$>$C mutation also has a larger proportion in \underline{A}N.

For the SNPs at the second position of the 3-mers (N\underline{C}N or N\underline{G}N)  as shown in \autoref{fig:3mer_2} (b), we observe the following mutation patterns.\\
(1) W\underline{G}N (where W is A or T) has C$>$T dominated mutation except for A\underline{G}G. \\
(2) S\underline{G}N (where S is G or C) has G$>$A dominated mutations. \\
(3) A\underline{G}G has high G$>$A mutations.\\
(4) Characteristic combinations S\underline{C}G (where S is G or C) are stable and only a few of G$>$T mutations are detected.\\
(5) Characteristic combinations G\underline{G}S (where S is G or C) are stable, having few G$>$T mutations.\\

For the SNPs at the third position of  3-mers (NN\underline{A} or NN\underline{T})  a shown in \autoref{fig:3mer_2} (a), we observe the following mutation patterns.\\
(1) A$>$G mutation has a high frequency in  NN\underline{A}.\\
(2) T$>$C mutation has a high frequency in  NN\underline{T}.\\
(3) T$>$C mutation is dominated in  NG\underline{T} and only a few of T$>$A and T$>$G are found in the sequence context of NG\underline{T}.\\

\begin{figure}[ht!]
    \centering
    \includegraphics[width=1\textwidth]{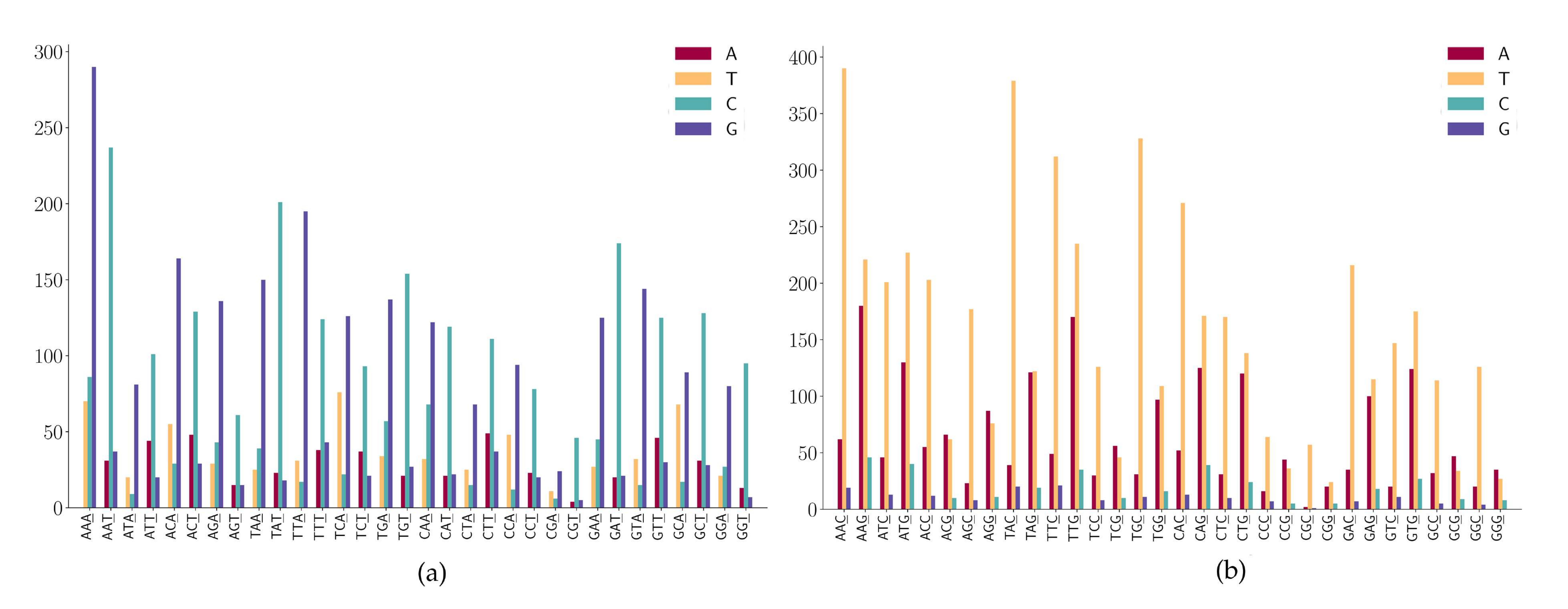}
    \caption{SNP frequency on the third positions of the 3-mer motifs.(a) A or T is on the third positions of 3-mer motifs. (b) C or G is on the third positions of 3-mer motifs.}
    \label{fig:3mer_3}
\end{figure}

For the SNPs at the third position of   3-mers (NN\underline{C} or NN\underline{G}) as shown in \autoref{fig:3mer_3} (b), we observe the following mutation patterns.\\
(1) NN\underline{C} has a high frequency in  C$>$T mutations.\\
(2) G$>$T mutation has a high frequency in   NN\underline{G}.\\
(3) G$>$A also highly expressed in the sequence context of NC\underline{G}.\\
(4) Characteristic combinations  CG\underline{C} are stable and the mutations on these patterns  are most likely to be C$>$T transitions.

\subsection{Coronavirus  evolution}

\begin{figure}[ht!]
    \centering
    \includegraphics[width=1\textwidth]{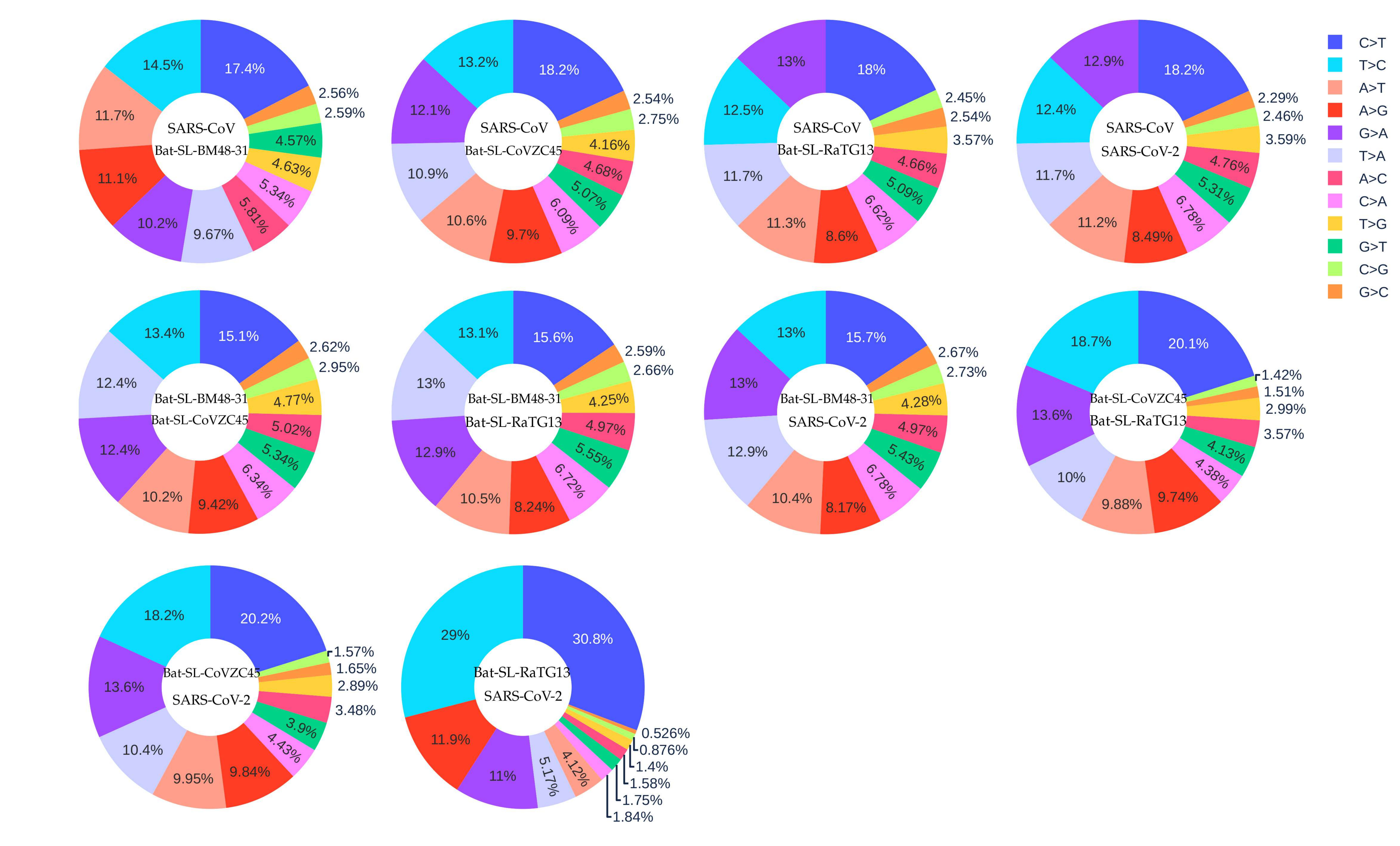}
    \caption{The distribution of 12 SNP types among SARS-CoV, Bat-SL-BM48-31, Bat-SL-CoVZC45, Bat-SL-RaTG13, and SARS-CoV-2. Here, the text on the top represents the reference genome and the text at the bottom represents the mutant sequence.}
    \label{fig:mutation}
\end{figure}

It is reasonable to assume that five coronaviruses  SARS-CoV (2003) \cite{lee2003major}, Bat-SL-BM48-31 (2008) \cite{drexler2010genomic}, Bat-SL-CoVZC45 (2017) \cite{hu2018genomic}, Bat-SL-RaTG13 (2013) \cite{zhou2020pneumonia}, and SARS-CoV-2 (2019) \cite{wu2020new} are of the same origin but differ from each other by their evolutionary stages. Among them, the data collection date of Bat-SL-RaTG13 (2013) was denoted as July 24, 2013 while the data was not uploaded to the GISIAD database until January 27, 2020.  
Figure   \ref{fig:mutation} shows the mutation ratio among these five genomes. First, similar to  SARS-CoV-2 mutations listed in  \autoref{tab:world ratio},  four transition types  (i.e., A$>$C,  C$>$A, C$>$T, and T$>$C) still have high mutation ratios. Particularly, C$>$T type has the highest ratio, indicating that host immune response still plays  the major role. However,  transversion type G$>$T is not as important as that in the SARS-CoV-2 mutations discussed early. Nonetheless, transversion types  A$>$T and T$>$A appear on the top six mutation types. 

We hypothesize that gene editing via APOBEC (C$>$T) and ADAR (A$>$G) is a driving force for RNA viral evolution as shown in  \autoref{tab:world ratio}. Viruses may fight back the host immune response with either defective repair or reversed mutations (T$>$C) within survived isolates. Therefore, T$>$C mutation rate would decrease during evolution.   We are interested in not only  the C$>$T transition ratio, but also the ratio of C$>$T over T$>$C, the reversed transitions.   From \autoref{tab:world ratio} and Figure   \ref{fig:mutation}, we can deduce that the following:\\ 
\begin{enumerate}
    \item From SARS-CoV-2 reference genome to 33693 genomes: C$>$T: 24.06\%, T$>$C: 14.53\% (Higher C$>$T ratio, relatively lower T$>$C ratio, and C$>$T to T$>$C ratio: 1.66)\\
    \item From SARS-CoV to Bat-SL-BM48-31:  C$>$T: 17.40\%, T$>$C: 14.50\% (Higher C$>$T ratio, relatively lower T$>$C ratio, and  C$>$T to T$>$C ratio: 1.20)\\
    \item From SARS-CoV to Bat-SL-CoVZC45:  C$>$T: 18.20\%, T$>$C: 13.20\% (Higher C$>$T ratio, relatively lower T$>$C ratio, and C$>$T to T$>$C ratio: 1.37)\\
    \item From SARS-CoV to Bat-SL-RaTG13:  C$>$T: 18.00\%, T$>$C: 12.50\% (Higher C$>$T ratio, relatively lower T$>$C ratio, and C$>$T to T$>$C ratio: 1.50)\\
    \item From SARS-CoV to SARS-CoV-2:  C$>$T: 18.20\%, T$>$C: 12.40\% (Higher C$>$T ratio, relatively lower T$>$C ratio, and C$>$T to T$>$C ratio: 1.47)\\
    \item From Bat-SL-BM48-31 to Bat-SL-CoVZC45:  C$>$T: 15.10\%, T$>$C: 13.40\% (Higher C$>$T ratio, relatively lower T$>$C ratio, and C$>$T to T$>$C ratio: 1.13)\\
    \item from Bat-SL-BM48-31 to Bat-SL-RaTG13:  C$>$T: 15.60\%, T$>$C: 13.10\% (Higher C$>$T ratio, relatively lower T$>$C ratio, C$>$T to T$>$C ratio: 1.19)\\
    \item From Bat-SL-BM48-31 to SARS-CoV-2:  C$>$T: 15.70\%, T$>$C: 13.00\% (Higher C$>$T ratio, relatively lower T$>$C ratio, and C$>$T to T$>$C ratio: 1.21)\\
    \item From Bat-SL-CoVZC45 to Bat-SL-RaTG13:  C$>$T: 20.10\%, T$>$C: 18.70\% (Higher C$>$T ratio, relatively lower T$>$C ratio, and C$>$T to T$>$C ratio: 1.07)\\
    \item From Bat-SL-CoVZC45 to SARS-CoV-2:  C$>$T: 20.20\%, T$>$C: 18.20\% (Higher C$>$T ratio, relatively lower T$>$C ratio, and C$>$T to T$>$C ratio: 1.11)\\
    \item From Bat-SL-RaTG13 to SARS-CoV-2:  C$>$T: 30.80\%, T$>$C: 29.00\% (Higher C$>$T ratio, relatively lower T$>$C ratio, and C$>$T to T$>$C ratio: 1.06)\\
\end{enumerate}
 
It is seen  that viral evolution order may be determined by the T$>$C over T$>$C ratio. By this analysis, we have the following evolution order for aforementioned coronaviruses, SARS-CoV (2003) $\to$ Bat-SL-BM48-31 (2008) $\to$ Bat-SL-CoVZC45 (2017) $\to$ Bat-SL-RaTG13 (2013) $\to$ SARS-CoV-2 (2019) $\to$ 33693 SARS-CoV-2 genome isolates (2020). Here, we have one reversed order between Bat-SL-CoVZC45 (2017) $\to$ Bat-SL-RaTG13 (2013).  This may happen for a few reasons. First, these coronaviruses may not be of the same origin. Second, the data collection date may not be accurate. The sequence of Bat-SL-RaTG13 (2013) was  not uploaded until 2020. Finally, our method may admit a few counterexamples.   
 
\section{Discussions}
The SNPs type distribution of 33693 SARS-CoV-2 isolates is listed in \autoref{tab:world ratio}. The C$>$T SNP mutation is remarkably higher than other mutation types. From the distribution of the 12 SNP types, we may infer that the excessive C$>$T transitions cannot explained by random mutations, instead, hypermutation C$>$T is due to the cytosine-to-uridine deamination gene editing in human host response.

\begin{figure}[ht!]
    \centering
    \includegraphics[width=1\textwidth]{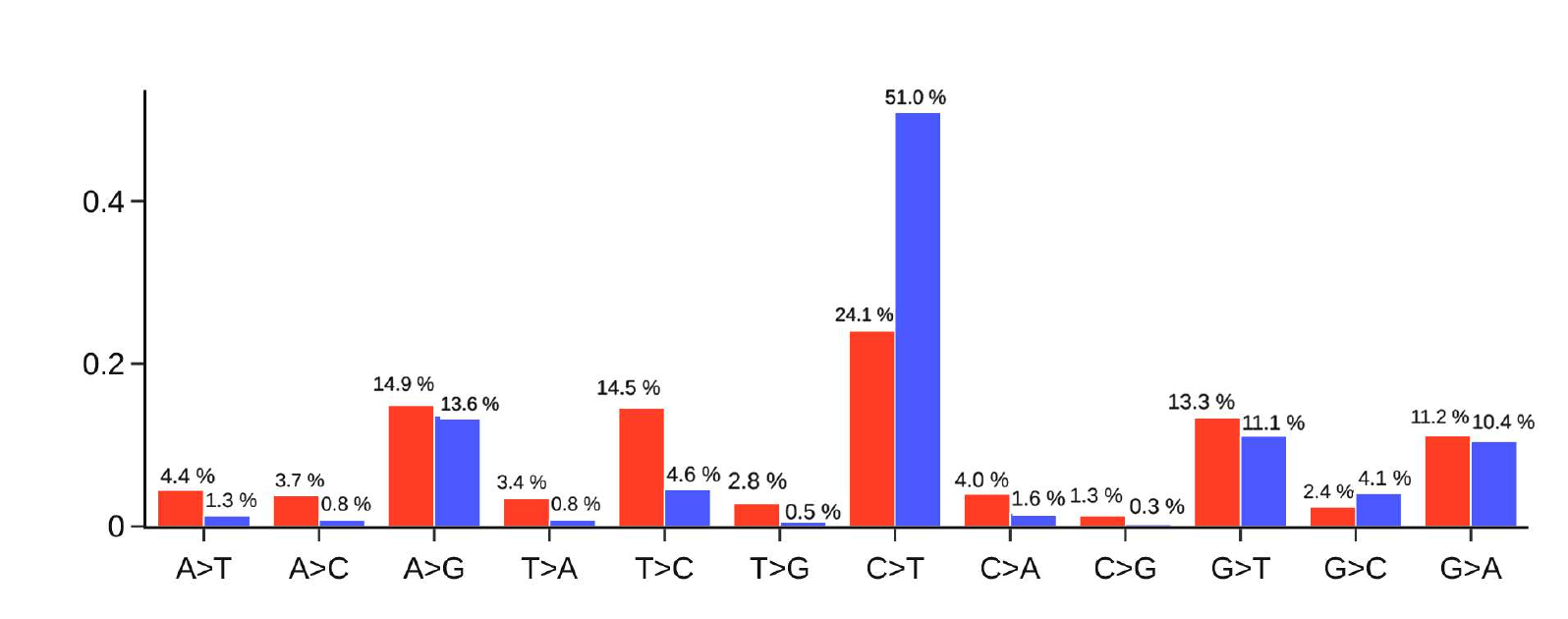}
    \caption{Comparison of the ratios of 12 SNP types among unique mutations (red) and non-unique mutations (blue) in SARS-CoV-2 genomes globally. Here, if we count the same mutation that appears in different SARS-CoV-2 isolates only once, then we call those mutations as unique mutations. If we count the same mutation in different SARS-CoV-2 isolates repeatedly according to their frequency, then all of the mutations that are detected in the complete SARS-CoV-2 genome sequences are called the non-unique mutations.}
    \label{fig:Single-duplicated-mutations_World}
\end{figure}

Figure \ref{fig:Single-duplicated-mutations_World} presents a comparison of the ratios of 12 SNP types of among unique and non-unique mutations over all of the SARS-CoV-2 genome isolates. The most striking feature is that the C$>$T  ratio is more than doubled in the non-unique mutations, which indicates the overwhelming host immune response to viral infection.   Another interesting feature is that the inverse transition T$>$C has a dramatic reduction of 68\% from the unique mutation ratio to the non-unique mutation ratio. These changes reflect the fact that many C$>$T mutations are high-frequency ones whereas virus reverses T$>$C mutations are of low frequency in nature.  The same explanation applies to many mutation types in Figure \ref{fig:Single-duplicated-mutations_World}  that have significantly reduced their ratios in the non-unique mutations.  However, we observed that ratios of mutation types A$>$G, G$>$T, and G$>$A do not change much in the non-unique mutations, reflecting the fact that these mutation types maintain a near-average frequency.  
 
Figure \ref{fig:Single-duplicated-mutations_World} shows that the second most frequent mutation type is A$>$G transitions, standing at 13.6 \%. The combined  C$>$T  and  A$>$G transition types account for near 65\% of all mutations. Therefore,  host gene editing via APOBEC and ADAR is the major driven force of SARS-CoV-2 evolution.

Neutralizing antibodies play a significant role in the clearance of viruses and have been considered a crucial immune artifact for the defense or treatment of viral diseases. However, a clinical study shows that five percent of people recovered from COVID-19 had no detectable antibodies \cite{wu2020neutralizing}. Another observation is that there are a large number of asymptomatic carrier transmission of {COVID-19} \cite{bai2020presumed}. The reason for the no-antibody COVID-19 recovers and asymptomatic carriers is unknown. From the mutation analysis in this study, the APOBEC3 RNA editing is implicated as a strong secondary defenses system for mutating virus, and consequently, mitigating infection. We postulate that COVID-19 recoveries or convalescents without antibody and some asymptomatic carriers are probably owing to the increased APOBEC3 activity in host immune systems.

\section{Methods and material}
\subsection{SNP genotyping}
Here, 33693 complete genomes of the SARS-CoV-2 strains of the globe are retrieved from the GISAID database \cite{shu2017gisaid} as of July 31, 2020. Only the complete genomes of high-coverage that have no stretches of 'NNNNN' include in the dataset. The complete genome sequences are aligned with the reference genome of SARS-CoV-2 by the MSA tool Clustal Omega using the default parameters \cite{sievers2014clustal}. The SNP mutations are retrieved from the aligned genomes according to the reference SARS-CoV-2 genome (GenBank access number: NC\_045512.2) \cite{wu2020new}. The SNP profile, including nucleotide changes and the corresponding positions in a genome, can be considered as the genotype of the virus.

\subsection{SNP analysis}
The Cluster Omega is employed to carry out the multiple sequence alignment. The genomic analytics is performed using computer programs in Python and Biopython libraries \cite{cock2009biopython}. 

\subsection{Data availability} The nucleotide sequences of the SARS-CoV-2 genomes used in this analysis are available, upon free registration, from the GISAID database (\url{https://www.gisaid.org/}). The SNP IDs and the acknowledgments of the SARS-COV-2 genomes are given in the Supporting Information.  

\section{Conclusion}
We use genotyping to analyze the mutation types and their distributions of SARA-CoV-2 genome isolates. 
We show that host gene editing, namely APOBEC (apolipoprotein B mRNA editing enzyme, catalytic polypeptide-like)  and  ADAR (adenosine deaminases acting on RNA), are the main driven forces of SARS-CoV-2 evolution, accounting for near 65\% recorded mutations.   
We reveal that the immune systems of children under age five and the elderly appear to overreact to SARS-CoV-2 infection and may be at high risk from COVID-19. Some minor gender dependence in immune response was also detected. 
We uncover that the populations of Oceania and Africa react significantly more intensive to SARS-CoV-2 infection than those of  Europe and Asia.
 Our study indicates that while systemic health and social inequities have put African Americans at increased risk of getting sick from COVID-19, their immune systems' overreacting to viral infection may have put them at increased risk of dying from COVID-19.
The mutational signatures have been analyzed to explore the preferred gene editing environments. 
Finally, we show that the ratio of mutation type C$>$T over T$>$C may be used to indicate the evolution direction and distinguish the evolution order between two genome sequences of the same origin. 

\section*{Supporting Information}
Supporting information is available for  supplementary figures, including the distribution of 12 SNP types among non-unique mutations, the distribution of 12 SNP types between each pair of 10 coronaviruses, and 
4-mer  analysis of mutational signatures.  Supplementary tables are available for GISAID IDs and GISAID acknowledgment.

\section*{Acknowledgment}
This work was supported in part by NIH grants  GM126189 and AI145504, NSF Grants DMS-1721024,  DMS-1761320, and IIS1900473,  Michigan Economic Development Corporation,  George Mason University award PD45722, Bristol-Myers Squibb, and Pfizer. The authors thank The IBM TJ Watson Research Center, The COVID-19 High Performance Computing Consortium, and  NVIDIA for computational assistance. 

 \clearpage
\bibliographystyle{unsrt}
 \bibliography{refs}
\end{document}